\documentclass{article}
\usepackage{graphicx} %
\usepackage[final]{mlncp_2023}
\usepackage{todonotes}
\usepackage{hyperref}
\usepackage{siunitx}
\DeclareSIUnit{\million}{\text{million}}
\usepackage{tikz}
 
\usepackage{booktabs}
\usepackage[normalem]{ulem}
\usepackage{float}
\usepackage[subtle]{savetrees}

\title{SpiNNaker2: A Large-Scale Neuromorphic System for Event-Based and Asynchronous Machine Learning}
\author{%
  Hector A. Gonzalez$^{1,2}$
  \;
  Jiaxin Huang$^2$
  \;
  Florian Kelber$^1$
  \;
  Khaleelulla Khan Nazeer$^1$
  \\
  \textbf{Tim Langer$^1$}
  \; 
  \textbf{Chen Liu$^1$}  
  \; 
  \textbf{Matthias Lohrmann$^{1,2}$}    
  \; 
  \textbf{Amirhossein Rostami$^1$}      
  \;  
  \textbf{Mark Sch\"one$^1$} 
  \;
   \\
  \textbf{Bernhard Vogginger$^1$}   
  \;
  \textbf{Timo C. Wunderlich$^3$}
  \; 
  \textbf{Yexin Yan$^1$}     
  \;   
  \textbf{Mahmoud Akl$^2$}       
  \;  
  \textbf{Christian Mayr$^1$}
  \\
  $^1$Technische Universit{\"a}t Dresden %
  \quad
  $^2$SpiNNcloud Systems GmbH
  \quad
  $^3$Universitätsmedizin Berlin
  \\
  \texttt{\{hector.gonzalez, jiaxin.huang, matthias.lohrmann, mahmoud.akl\}@spinncloud.com,} \\
  \texttt{\{florian.kelber, khaleelulla.khan\_nazeer, tim.langer, chen.liu,} \\ \texttt{amirhossein.rostami, mark.schoene, bernhard.vogginger, yexin.yan,} \\ 
  \texttt{christian.mayr\}@tu-dresden.de}, \texttt{timo.wunderlich@bih-charite.de} \\
}

\begin{document}

\maketitle

\begin{abstract}
The joint progress of artificial neural networks (ANNs) and domain specific hardware accelerators such as GPUs and TPUs took over many domains of machine learning research.
This development is accompanied by a rapid growth of the required computational demands for larger models and more data.
Concurrently, emerging properties of foundation models such as in-context learning drive new opportunities for machine learning applications.
However, the computational cost of such applications is a limiting factor of the technology in data centers, and more importantly in mobile devices and edge systems.
To mediate the energy footprint and non-trivial latency of contemporary systems, 
neuromorphic computing systems deeply integrate computational principles of neurobiological systems by leveraging low-power analog and digital technologies.
SpiNNaker2 is a digital neuromorphic chip developed for scalable machine learning.
The event-based and asynchronous design of SpiNNaker2 allows the composition of large-scale systems involving thousands of chips. This work features the operating principles of SpiNNaker2 systems, outlining the prototype of novel machine learning applications. These applications range from ANNs over bio-inspired spiking neural networks to generalized event-based neural networks. 
With the successful development and deployment of SpiNNaker2, we aim to facilitate the advancement of event-based and asynchronous algorithms for future generations of machine learning systems.
\end{abstract}

\section{Introduction}
\label{sec:introduction}
Progress in machine learning and the availability of computational resources are tightly coupled. Especially, the breakthrough of deep learning can be attributed to the successful acceleration of deep neural network (DNN) training on large-scale data with graphics processing units (GPUs) \citep{Krizhevsky2012}.
Deep learning models continue to improve their task performance with the number of floating-point operations spent on training on a wide variety of tasks.
Therefore, deep learning models have been scaled up at unprecedented speed up to the limit of compute availability \citep{Sevilla2022, thompson2022computational}.
How far can we take the joint scaling of model parameters and hardware accelerators?
Since the 90s, the peak hardware floating-point operations per second (FLOPS) grew by an average rate of \num{60000} per \num{20} years.
However, the DRAM bandwidth and interconnect bandwidth only grew by an average rate of \num{100} and \num{30} per \num{20} years, respectively \cite{gholami2020}. 
This development changes the requirements on algorithms and accelerators, with the focus shifting to a reduction of the overall communication in future machine learning systems.

Considering the highly scalable computational substrate of biological nervous systems, we observe that communication in the brain is 
a) locally dense but globally very sparse,  
b) temporally very sparse via binary spike communication, and
c) asynchronous.
The field of neuromorphic computing develops algorithms and accelerators that leverage these principles to devise efficient and scalable systems.
The core algorithmic concept to deliver these goals are spiking neural networks (SNNs).
Although, many works promise high savings in energy consumption, achieving state-of-the-art performance on machine learning benchmarks proves to be difficult with SNNs \citep{Tavanaei2019, Taherkhani2020, Yamazaki2022}.
Recent works aim to bridge the performance gap between SNNs and DNNs.
In particular, \cite{Wozniak2020} and \cite{Subramoney2023} combine deep neural network architectures with discontinuous step functions and state resets to avoid communication in their networks. 
We are in favor of broader research efforts on communication avoiding learning systems.

While accelerators propel deep learning research forward, this research in turn channels unprecedented investments into the development of accelerators tailored to the needs of the deep learning community. Such interplay between deep learning and accelerators creates a path dependency: The performance of dense DNNs drives investments in dense accelerators, which subsequently boosts the performance of dense DNNs, leading to further investment in these accelerators \citep{Barham2019, Hooker2021}. This dynamic prompts the question of whether alternative paths exist. Are there fundamentally different combinations of algorithms and hardware that yield more efficient learning machines, assuming they benefit from the same technological advancements that dense DNNs and GPUs have experienced over the past decades?

To address the previous challenges, we present SpiNNaker2, a versatile accelerator for event-based and asynchronous machine learning (ML).
SpiNNaker2 is a highly parallel system composed of asynchronously operating processing elements (PEs) interconnected by an efficient network on chip \citep{Hoeppner2022}.
Model designers working with SpiNNaker2 are neither limited by highly structured thread execution (compared to GPUs), constrained by procedural architectures lacking event-based awareness (e.g., pure DNN platforms \citep{sambanovaISSCC2022,tenstorrentISSCC2022}) nor do they have to restrict themselves to specific neuron implementations (compared to pure neuromorphic systems \citep{davies2018loihi,pehle2022brainscales}).
More than 35k SpiNNaker2 chips were manufactured and are assembled into the world's largest brain-like supercomputer with about 5~million processing elements, which will be remotely available to interested researchers around the world. %
With this workshop contribution, we aim to accelerate the exploration of event-based and synchronous machine learning models as an alternative path to GPU-centric models.

\section{The SpiNNaker2 System}
\label{sec:s2-system}
\begin{figure}[!t]
    \centering
    \includegraphics[width=\textwidth ]{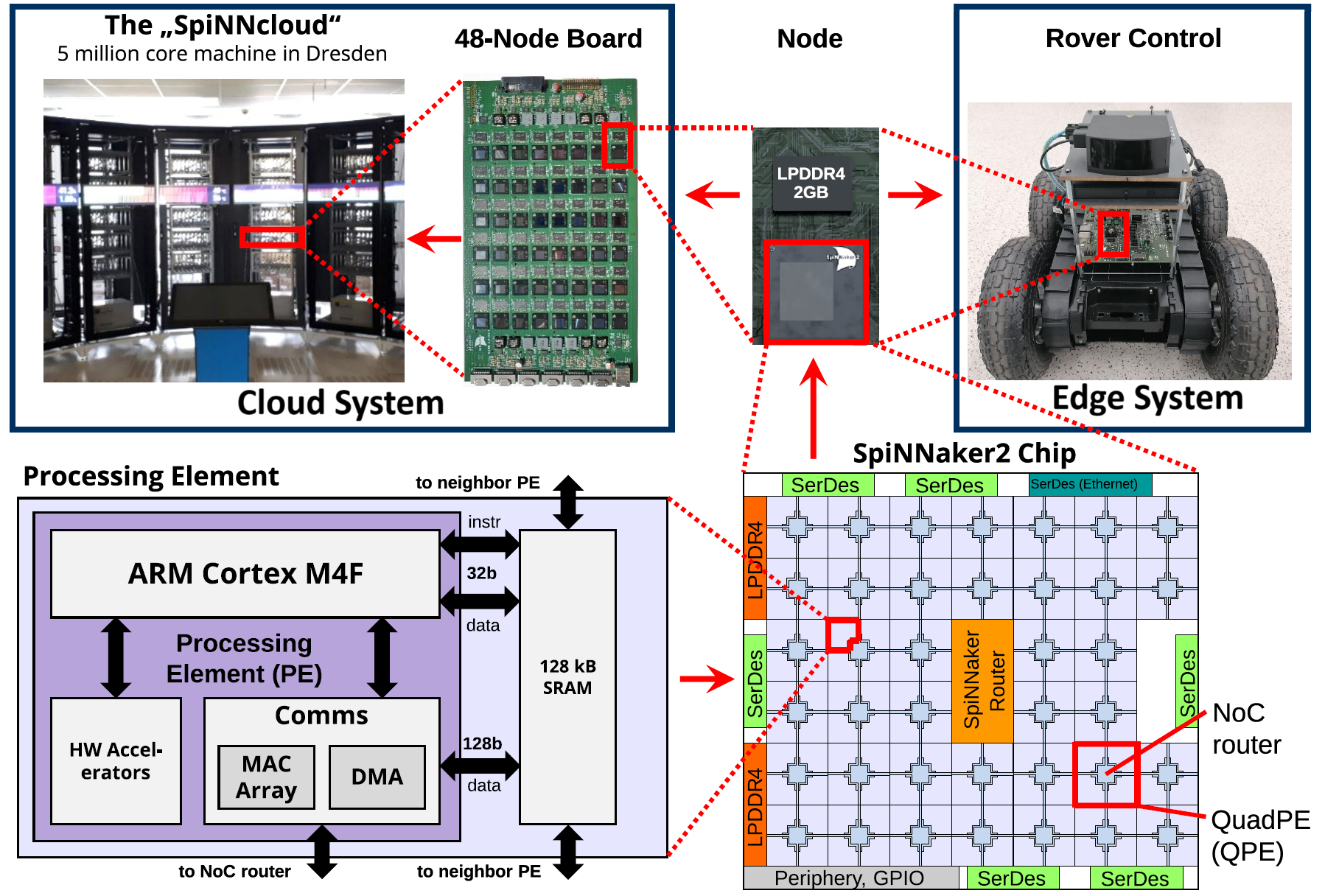}
    \caption{Overview of SpiNNaker2 architecture for the "SpiNNcloud" cloud system and edge systems.}
    \label{fig:spinnaker2}
\end{figure}

SpiNNaker2 is a massively parallel compute system that can be scaled up from one standalone chip with 152 ARM Cortex M4F cores (e.g., complex edge systems as in Fig.~\ref{fig:spinnaker2}) to millions of cores (e.g., supercomputer levels as in Fig.~\ref{fig:spinnaker2}). Compared to conventional multiprocessing systems, there is no operating system that  dynamically schedules compute tasks to cores with shared virtual memory. Instead, each core runs a small pre-compiled program on 128kB SRAM that executes simple tasks upon reception of \emph{events}.
Technically, the event-based processing is handled through ARM's interrupt controller that starts, pauses and resumes user functions depending on interrupts (IRQs), while utilizing a core sleep mode to save energy.

There are different means on how applications on different cores can communicate with each other.
Within a SpiNNaker2 chip, the Network-on-Chip (NoC) offers high-speed point-to-point communication between cores and access to the off-chip DRAM. DMA (direct memory access) units in each core and the DRAM interfaces enable bulk transfer without stopping the cores.
For scalable, system-wide communication, each chip has a dedicated SpiNNaker2 packet router, containing configurable routing tables, and 6 links to neighbour chips.
On SpiNNaker2, different packet types with up to 128-bit payload allow the efficient communication between PEs, chips and boards. As an example, in SNN simulation multi-cast packets are used for the transmission of spikes where a 32-bit key represents the ID of the spiking neuron.

As the SpiNNaker2 system aims to speed up SNNs, DNNs or hybrid approaches, each core provides selected operation acceleration. These include exponential, logarithm, true and pseudo random number generation, as well as energy-efficient low-precision 8-bit/16-bit integer matrix multiplication and 2-dimensional convolution. The event-based nature of the system encourages autonomous and asynchronous execution between cores. We leverage this by dynamically adapting clock frequency 
and supply voltage of individual cores to further save energy. This power switching can be automated by being coupled to the application \citep{Hoeppner2019,  Hoeppner2022,Hoeppner2017,Yexin2022}. Each chip has 2GB of DRAM to form a node and support memory-intensive DNN execution.
PCBs with 48 of those nodes are used to build the 5-million core large-scale system. The chip connectivity in those boards rely on a hexagonal grid, assembling a torus-shaped network at a system level, which reduces the number of node hops compared to mesh interconnects.

The main advantages of the SpiNNaker2-based systems then include its event-based nature, its native support for hybrid models beyond pure DNNs or SNNs (e.g., including symbolic models \citep{hammer2022reasoning}), and its high-speed infrastructure to scale toward arbitrary large systems without losing real-time capabilities. However, programming and operating those massively parallel systems is challenging.
Compute problems such as SNNs, DNNs, graphical models, or sensor processing pipelines require small task partitioning, mapping, setting up event-based communication, data loading and result readings.
In the following section we will show operating examples on single-chip SpiNNaker2 systems. Larger applications are planned via the operating principle from SpiNNaker1 (\citep{brown2014spinnaker,rowley2019spinntools}).

\section{Applications}
\label{sec:s2-apps}
The first generation SpiNNaker system \citep{SpiNNaker1, SpiNNaker1system} is used in 23 countries by more than 60 research groups ranging 
from neuromorphic computing to neuroscience \citep{spinnworld,SpiNNaker2020book}.
The SpiNNaker2 project expands the scope of algorithmic research to cover models from neuromorphic computing up to conventional machine learning models. In the following section, we present how 
ANNs
and SNNs are mapped to SpiNNaker2, concluding
with our initial work on event-based ANNs for both inference and training. Such
hybrid approach takes the best of both worlds: The numerical simplicity and scalability of ANNs, combined with the ability of biology to reduce data flow and computation to the minimum necessary for a given task. 

\subsection{Artificial Neural Networks}
A major use case of the SpiNNaker2 platform is the energy-efficient training and inference of ANNs, for which two approaches will be presented in sections \ref{sec:scheduling} and \ref{sec:deep-rewiring}.

\label{sec:dnns}

\begin{figure}[!t]
    \centering
    \centerline{\includegraphics[width=1.0\linewidth]{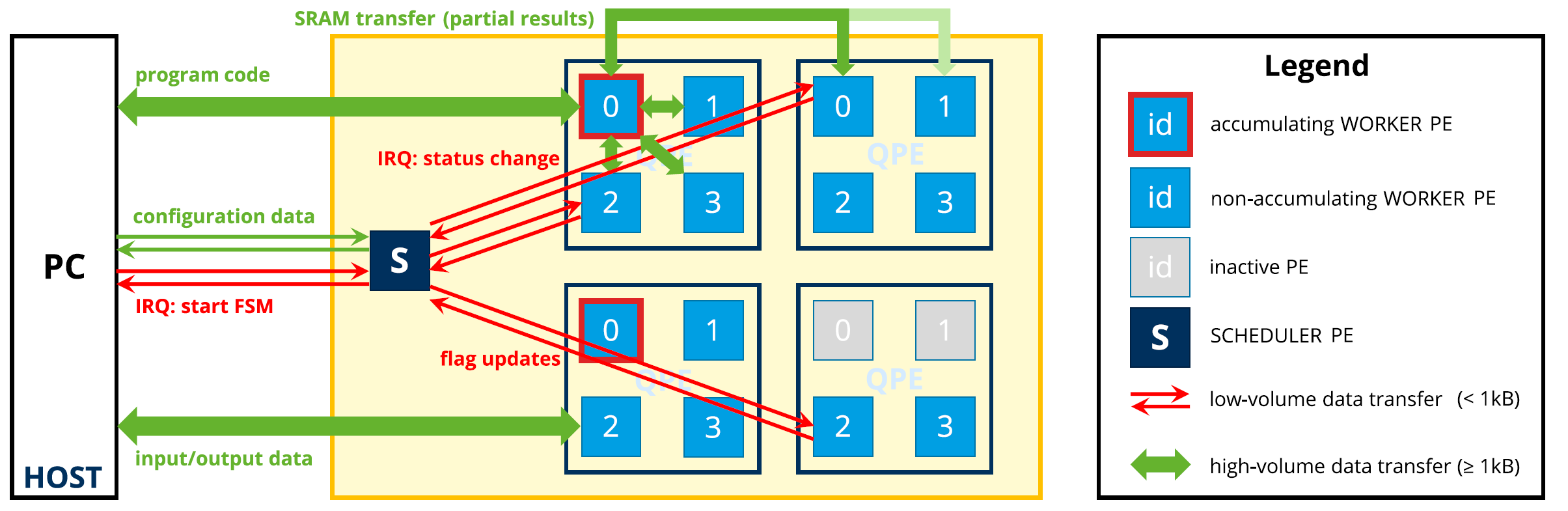}}
    \caption{Control (red) and Data (green) flow of the graph execution with a host, \textit{scheduler} and \textit{worker} PEs.}
    \label{fig:scheduling_approach}
\end{figure}

\subsubsection{Scheduling Operations for ANNs}
\label{sec:scheduling}
Considering the large dimensions of current ANN architectures, the mapping of ANN layers onto PEs with limited storage and local accelerators becomes a challenge. Therefore, a scheduling approach for SpiNNaker2 has been developed to distribute and efficiently compute layers across a multitude of PEs. 
The current scheduling approach (see Fig.~\ref{fig:scheduling_approach}) defines a single \textit{scheduler} that coordinates the state transitions of \textit{workers} using a finite state machine (FSM). After an initial interrupt request (IRQ) by the \textit{host}, the asynchronous schedule execution is completely orchestrated on-chip by the \textit{scheduler} PE to avoid slow chip-to-host communication via Ethernet. The schedule is executed statically (i.e.  assignment of tasks to PEs and task execution sequence are pre-defined a-priori by \textit{host}). The data exchange is organized decentrally and locally between PEs to avoid communication bottlenecks via the \textit{scheduler}, while the control flow of status words is organized centralistically and hierarchically from the \textit{scheduler} to the \textit{workers} (see Fig.~\ref{fig:scheduling_approach}).

As an example, a distributed matrix-matrix multiplication requires the tiling of each input matrix into submatrices, the calculation of partial results by multiplication of the respective submatrices, the transfer of these partial results, and the accumulation of the corresponding partial results. The tiled input matrices will be transferred from the \textit{host} to the \textit{workers}. After an initial IRQ by the \textit{scheduler}, the \textit{workers} will start executing the multiplication of the submatrices. Having finished, each \textit{worker} will write its updated flags to the \textit{scheduler's} SRAM. In a loop, the \textit{scheduler} is checking whether the state of any \textit{worker} PE changed, and if positive, perform a state update. If a \textit{worker} finished the matrix multiplication, it will either terminate, or fetch and locally accumulate the partial results of another \textit{worker} within a pre-defined accumulation group. After the status update, the \textit{scheduler} will send an IRQ towards the accumulating \textit{worker} to initiate the fetching from another \textit{worker's} SRAM. Once the updated operation flags are sent 
to the \textit{scheduler}, the FSM update loop registers a flag change, 
updating
the internal state of the specific \textit{worker}, and 
triggering
an IRQ combined with a flag update to the \textit{worker}.

Future scheduling approaches will tailor full state-of-the-art ANNs such as large language models (LLMs) into the supercomputer fabric to reduce power consumption via the energy-proportional SpiNNaker2 features (e.g., event-based). Simulation frameworks such as \cite{Kelber2020} will be extended, including other \textit{scheduler-worker} interactions (e.g., subgroup scheduling), as well as dynamic scheduling strategies.

\subsubsection{Sparse-to-Sparse Training: Deep Rewiring}
\label{sec:deep-rewiring}
The sampling behavior observed in biology \cite{Kappel2018,Yan2019} demonstrates memory-efficient learning. Analogously, a learning algorithm known as deep rewiring maintains a consistent level of sparsity by dynamically disconnecting insignificant synapses and randomly establishing connections elsewhere throughout the entire training process. This approach facilitates learning in memory-constrained environments, particularly on edge devices.

The application of deep rewiring to SpiNNaker2 prototype chip \cite{Liu2018} showcases a highly sparse stochastic gradient descent (SGD) training from scratch. Remarkably, it achieves a 96.6\% accuracy on the MNIST dataset for handwritten digits with 3 dense layers involving 410 neurons, while operating within a tight memory constraint of 64 kB and maintaining a connectivity of 1.3\%. Time profiling reveals that the rewiring step dominates the computation time. This impact is mitigated by adjusting the rewiring frequency across iterations and leveraging the exponential accelerator and random number generator integrated in SpiNNaker2.
The training time of the 4-core deep rewiring on SpiNNaker2 is on par with that of a standard X86 CPU (Intel i5-6500), while the energy consumption is substantially reduced by two orders of magnitude. These findings underscore the potential of incorporating bio-inspired learning algorithms, such as deep rewiring, into SpiNNaker2, signaling a paradigm shift towards a more energy-efficient computation.

\subsection{Spiking Neural Networks}
\label{sec:snns}

Despite SpiNNaker2 innovating with event-based ANNs, its performance in bio-inspired SNNs remains being unique as it operates at scales that are not reachable by other neuromorphic systems. 
SNNs have the potential to be more efficient than 
conventional 
ANNs
because they mimic the brain mechanism of sparse communication between neurons and only transmit spikes when necessary, which greatly reduces the energy footprint. In general, a series of transformation steps, such as parsing and preprocessing SNN model information, are required to build the bridge between the trained SNN model and inference execution on SpiNNaker2.
The transformation steps start from the specifically trained SNN model (see section~\ref{sec:event-based}) or from an ANN-converted SNN model (e.g. using SNN toolbox \cite{ref_snntoolbox}). Then the SNN model is interpreted into an application graph, presenting the spike flow path among neuron populations. Each population is further split into one or several sub-populations to fit the SRAM resource of each PE. All the sub-populations and the corresponding projections form a machine graph. The connection relations of these sub-populations contribute to the generation of a routing table. Finally, these transformed results are presented with 
c
files
to be
loaded on SpiNNaker2 along with an input spike train before execution. This mapping framework enables large-scale SNN simulation on multiple PEs of SpiNNaker2.

During runtime, the event-based synapse processing, time-triggered neuron state update and spike-based communication are applied to SNN execution. An event is a spike from a pre-synaptic neuron of another PE, and this spike triggers the synapses in the current PE to start processing it. The neuron states of the current PE are updated by programmable neuron models at a predefined regular time interval. Then the generated spikes are sent to the PEs with the post-synaptic neurons. This process has been showcased with synfire chain model, bursting network, an asynchronous irregular firing network from \cite{Hoeppner2019} and radar gesture recognition demonstration from \cite{ref_jennifer_aicas22_1} \cite{ref_jennifer_aicas22_2} as examples. Besides, such process can be accelerated 
by exploiting the MAC array on SpiNNaker2, with the spatial-temporal performance improved to some extent \cite{ref_jennifer_AICAS23} \cite{ref_jennifer_frontiers}. %
Very recently, the neuromorphic intermediate representation (NIR, \cite{pedersen2023neuromorphic}) was introduced offering an exchange format for SNN. Currently NIR is supported by 7 neuromorphic simulators and 4 hardware platforms. It allows to train deep SNN in frameworks like snnTorch \cite{eshraghian2021} or Norse \citep{norse2021} and deploy them on SpiNNaker2 using py-spinnaker2 \cite{vogginger2023pyspinnaker2}.

\subsection{Event-Based Artificial Neural Networks}
\label{sec:event-based}

\subsubsection{Event-Based Gated Recurrent Unit}
\label{sec:egru}
\begin{figure}[!t]
  \begin{minipage}{\linewidth}
      \centering
      \begin{minipage}{0.45\linewidth}
          \begin{figure}[H]
              \centering
                \includegraphics[width=\linewidth]{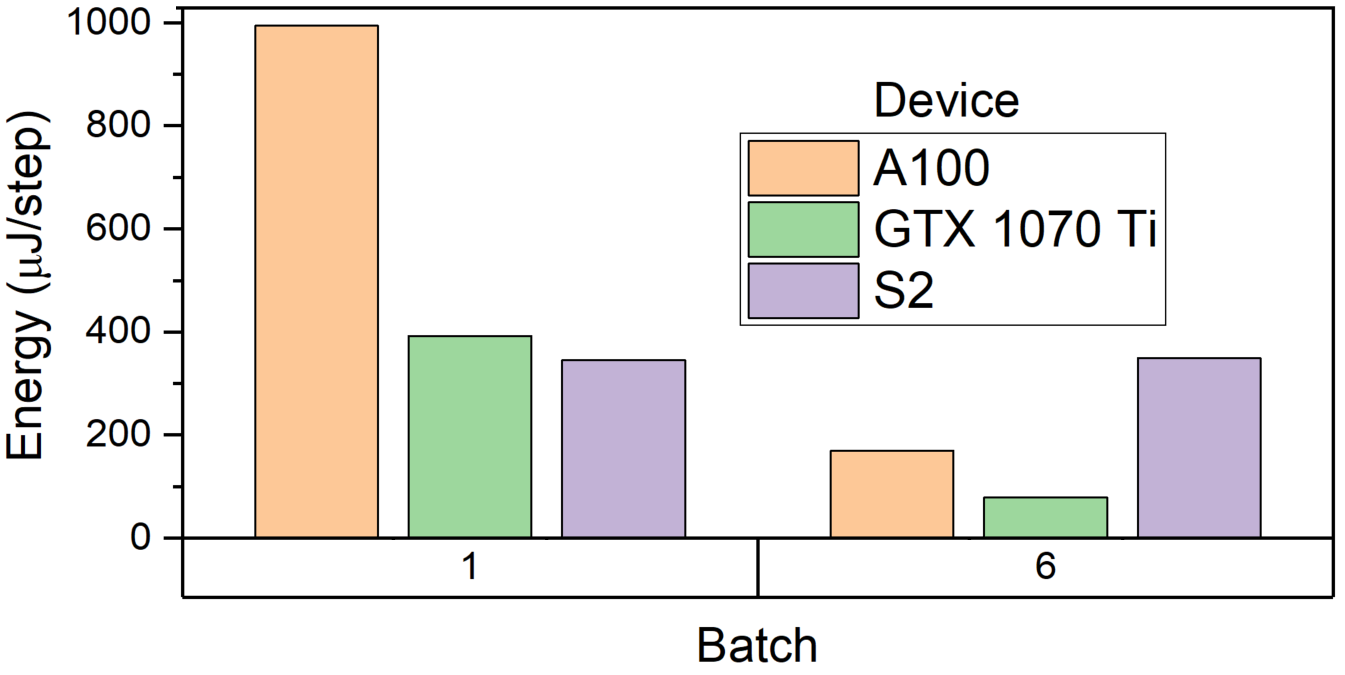}
                \caption{Energy per timestep for EGRU.
                }
                \label{fig:egru-energy}
          \end{figure}
      \end{minipage}
      \hspace{0.05\linewidth}
      \begin{minipage}{0.45\linewidth}
          \begin{figure}[H]
    \centering
    \includegraphics[width=\linewidth]{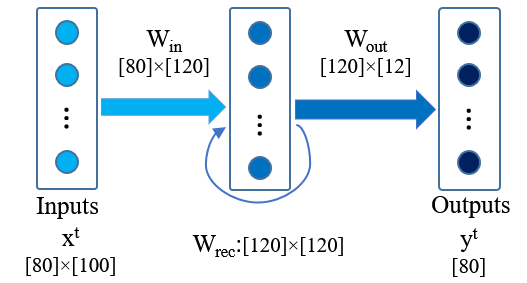}
    \caption{SRNN used in the e-prop algorithm.}   
    \label{fig:eprop-sch}
\end{figure}
      \end{minipage}
  \end{minipage}
\end{figure}

\citet{Subramoney2023} propose the Event-based Gated Recurrent Unit (EGRU) to combine the energy efficiency of SNNs and the performance of ANNs.
Therefore, EGRU employs a bio-inspired activity sparsity mechanism that allows the units to emit discrete and sparse-in-time events for communication.
Since events are sent sparingly, this leads to substantial computational savings during training and inference. To validate these claims, 
a 2 layer EGRU model
is implemented
and distributed it over 128 PEs on a single SpiNNaker2 chip.
The model is trained on a GPU similar to \citet{Subramoney2023} on the DVS gesture recognition task \citep{amirDVSdataset} and the weights are transferred to the SpiNNaker2 PE memory.
A CNN head is used to preprocess the dataset that is stored on the DRAM to be loaded onto local memory for every sample.
Fig. ~\ref{fig:egru-energy} shows the energy-per-timestep measurements,
 normalized over 18 samples of DVS gestures. 
The comparison is made with 
inference on 2 GPUs (Nvidia A100 \& GTX1070Ti), noticing a %
lower energy consumption on SpiNNaker2 compared to GPU inference. SpiNNaker2 displays a remarkable performance at batch-one or real-time conditions, but its energy remains constant for larger batches.
In terms of EGRU training,
\citet{Subramoney2023} also shows an event-based learning rule similar to EventProp in the limit of continuous time.
The cell state equation of the GRU can be viewed as the Euler discretization of a continuous time dynamical system.
Based on the theory of adjoint sensitivity analysis 
in EventProp
(see section~\ref{sec:event-prop}), \citet{Subramoney2023} derives the adjoint equations for a GRU system 
under
discrete state transitions triggered by input events.
Such a system is similarly suited for a SpiNNaker2 implementation.

\subsubsection{EventProp: Event-Based Backpropagation}
\label{sec:event-prop}
EventProp \cite{Wunderlich2021} is a learning algorithm for event-based backpropagation that computes exact gradients in 
SNNs
while retaining the temporal sparsity of spike-based communication during the backward pass. 
In a \emph{proof-of-concept} demonstration, we show that SpiNNaker2 can implement EventProp for multi-layer feed-forward SNNs.
Every SpiNNaker2 core implements a clock-driven simulation of a layer of leaky integrate-and-fire neurons.
During the forward pass, spike events are distributed across cores using multi-cast packets routed by the network-on-chip.
During the backward pass, spike events are distributed in reverse and carry error signals contained in the 128-bit payload of the multi-cast packets.
After the backward pass, a designated control core collects gradients from cores and computes weight updates using the Adam optimizer \cite{KingBa15}.
This allows for batch-parallel processing by accumulating gradients from different copies of the same layer, computed
on different cores.
The viability of this approach is demonstrated with a time-to-first-spike loss function, a single hidden layer network, and by achieving latency-encoded parallel training.
The results show that SpiNNaker2 supports event-based backpropagation through multi-cast event routing, along with batch execution.

\subsubsection{E-prop: Biologically-Inspired Learning}
\label{sec:e-prop}

Eligibility propagation (e-prop), a biologically-inspired online learning rule for Spiking Recurrent Neural Networks (SRNNs), serves as an alternative to Back Propagation Through Time (BPTT) \citep{bellecSolutionLearningDilemma2020}. The e-prop gradient at any given time step is independent of subsequent step information, enhancing memory efficiency and suitability for online learning. As validation, a 3-layer neural network (Fig.~\ref{fig:eprop-sch}) classifying the 12-category Google Speech Command dataset was implemented in \cite{AmirEprop}. While BPTT required 859KB of memory, e-prop used only 682KB (20\% less). For parallelization, synapses were evenly distributed across the PEs, allowing for local gradient computation at each time step and requiring only the spike transmission. Nevertheless, in the last time step, the output error is broadcast to other PEs for weight updating. The results show that SpiNNaker2 enables efficient spike transmission in algorithms such as e-prop, along with high test accuracies (i.e., 91.12\%) under real-time (batch-one) conditions.

\section{Discussion and Outlook}
\label{sec:outlook}
The proposed SpiNNaker2 system
enables large-scale simulation of event-based and asynchronous machine learning systems, two essential properties of scalable computational systems. %
The system deviates from the major direction focused by the deep learning community with the perspective to allow gains in energy efficiency and latency at scale. Turning from mostly dense and synchronous processing to event-based and asynchronous processing sets significantly different requirements for the development of learning algorithms. Despite the programming challenges, SpiNNaker2 paves the path for future fully event-based supercomputers with a neglectable Amdahl limit and an energy-proportional operation beyond DNNs. 

Among the projects described in this paper, EventProp and the event-based learning rule for EGRU demonstrate error propagation in event-based and potentially asynchronous neural networks. %
E-prop shows the utilization of locally restricted error signals to save communication and hence energy during training. In addition, sparse-to-sparse training in the Deep Rewiring approach shows how to improve further the efficiency. 

With the aim of driving research in this algorithmic front, the 5-million core system in Dresden, Germany, will 
grant access to
researchers keen on exploring these %
challenges with us. Such supercomputer finds applications in energy-efficient LLMs by leveraging for example event-based recurrent structures in the recent resurgence of RNNs against transformers at language modeling tasks \citep{dao2022hungry,peng2023rwkv, sun2023retentive}. Furthermore, the applications also include but are not limited to the new development of event-based ML, the large-scale deployment of hybrid models such as NARS \citep{hammer2022reasoning}, complex brain simulation, the utilization of the massive parallelism in probabilistic computing and distributed drug discovery. %
Systems with arbitrary sizes are also commercially available in \citep{SpiNNcloudSystems}.

\section*{Acknowledgements}
MS is fully funded by the Bosch Research Foundation.
The authors gratefully acknowledge the GWK support for funding this project by providing computing time through the Center for Information Services and HPC (ZIH) at TU Dresden. The authors also acknowledge the EIC Transition under the "SpiNNode" project (grant number 101112987), and the support by the German BMBF (Joint project 6G-life, ID: 16KISK001K and DAAD project SECAI, ID: 57616814).
This work was partially funded by the German Federal Ministry for Economic Affairs and Climate Action (BMWK) under the contract 01MN23004F.
Partially funded by the German Research Foundation (DFG, Deutsche Forschungsgemeinschaft) as part of Germany’s Excellence Strategy – EXC 2050/1 – Project ID 390696704 – Cluster of Excellence “Centre for Tactile Internet with Human-in-the-Loop” (CeTI) of Technische Universität Dresden.

\bibliography{references}

\begin{thebibliography}{48}
\providecommand{\natexlab}[1]{#1}
\providecommand{\url}[1]{\texttt{#1}}
\expandafter\ifx\csname urlstyle\endcsname\relax
  \providecommand{\doi}[1]{doi: #1}\else
  \providecommand{\doi}{doi: \begingroup \urlstyle{rm}\Url}\fi

\bibitem[Spi(2023)]{SpiNNcloudSystems}
{SpiNNcloud Systems GmbH}, 2023.
\newblock URL \url{https://spinncloud.com/}.
\newblock Accessed on October 3, 2023.

\bibitem[spi(2023)]{spinnworld}
{SpiNNworld: SpiNNaker presence worldwide}, 2023.
\newblock URL
  \url{https://www.google.com/maps/d/u/0/edit?mid=1jrbV2OVaBFqGlVYMxerSh0Pcexd_wznQ&ll=1.5170674512027844%2C0&z=2}.
\newblock Accessed on October 3, 2023.

\bibitem[Amir et~al.()Amir, Taba, Berg, Melano, McKinstry, Nolfo, Nayak,
  Andreopoulos, Garreau, Mendoza, Kusnitz, Debole, Esser, Delbruck, Flickner,
  and Modha]{amirDVSdataset}
Arnon Amir, Brian Taba, David Berg, Timothy Melano, Jeffrey McKinstry,
  Carmelo~Di Nolfo, Tapan Nayak, Alexander Andreopoulos, Guillaume Garreau,
  Marcela Mendoza, Jeff Kusnitz, Michael Debole, Steve Esser, Tobi Delbruck,
  Myron Flickner, and Dharmendra Modha.
\newblock A {{Low Power}}, {{Fully Event-Based Gesture Recognition System}}.
\newblock page~10.

\bibitem[Barham and Isard(2019)]{Barham2019}
Paul Barham and Michael Isard.
\newblock Machine learning systems are stuck in a rut.
\newblock In \emph{Proceedings of the Workshop on Hot Topics in Operating
  Systems}, HotOS '19, page 177–183, New York, NY, USA, 2019. Association for
  Computing Machinery.
\newblock ISBN 9781450367271.
\newblock \doi{10.1145/3317550.3321441}.
\newblock URL \url{https://doi.org/10.1145/3317550.3321441}.

\bibitem[Bellec et~al.(2020)Bellec, Scherr, Subramoney, Hajek, Salaj,
  Legenstein, and Maass]{bellecSolutionLearningDilemma2020}
Guillaume Bellec, Franz Scherr, Anand Subramoney, Elias Hajek, Darjan Salaj,
  Robert Legenstein, and Wolfgang Maass.
\newblock A solution to the learning dilemma for recurrent networks of spiking
  neurons.
\newblock \emph{Nature Communications}, 11\penalty0 (1):\penalty0 3625, July
  2020.
\newblock ISSN 2041-1723.
\newblock \doi{10.1038/s41467-020-17236-y}.

\bibitem[Brown et~al.(2014)Brown, Furber, Reeve, Garside, Dugan, Plana, and
  Temple]{brown2014spinnaker}
Andrew~D Brown, Steve~B Furber, Jeffrey~S Reeve, Jim~D Garside, Kier~J Dugan,
  Luis~A Plana, and Steve Temple.
\newblock Spinnaker—programming model.
\newblock \emph{IEEE Transactions on Computers}, 64\penalty0 (6):\penalty0
  1769--1782, 2014.

\bibitem[Dao et~al.(2022)Dao, Fu, Saab, Thomas, Rudra, and
  R{\'e}]{dao2022hungry}
Tri Dao, Daniel~Y Fu, Khaled~K Saab, Armin~W Thomas, Atri Rudra, and
  Christopher R{\'e}.
\newblock Hungry hungry hippos: Towards language modeling with state space
  models.
\newblock \emph{arXiv preprint arXiv:2212.14052}, 2022.

\bibitem[Davies et~al.(2018)Davies, Srinivasa, Lin, Chinya, Cao, Choday, Dimou,
  Joshi, Imam, Jain, et~al.]{davies2018loihi}
Mike Davies, Narayan Srinivasa, Tsung-Han Lin, Gautham Chinya, Yongqiang Cao,
  Sri~Harsha Choday, Georgios Dimou, Prasad Joshi, Nabil Imam, Shweta Jain,
  et~al.
\newblock Loihi: A neuromorphic manycore processor with on-chip learning.
\newblock \emph{Ieee Micro}, 38\penalty0 (1):\penalty0 82--99, 2018.

\bibitem[Eshraghian et~al.(2023)Eshraghian, Ward, Neftci, Wang, Lenz, Dwivedi,
  Bennamoun, Jeong, and Lu]{eshraghian2021}
Jason~K Eshraghian, Max Ward, Emre Neftci, Xinxin Wang, Gregor Lenz, Girish
  Dwivedi, Mohammed Bennamoun, Doo~Seok Jeong, and Wei~D Lu.
\newblock Training spiking neural networks using lessons from deep learning.
\newblock \emph{Proceedings of the IEEE}, 111\penalty0 (9):\penalty0
  1016--1054, 2023.

\bibitem[Furber and Bogdan(2020)]{SpiNNaker2020book}
Steve Furber and Petrut Bogdan, editors.
\newblock \emph{{SpiNNaker: A Spiking Neural Network Architecture}}.
\newblock now publishers, Boston-Delft, 2020.
\newblock \doi{10.1561/9781680836523}.
\newblock URL \url{http://dx.doi.org/10.1561/9781680836523}.

\bibitem[Furber et~al.(2014)Furber, Galluppi, Temple, and
  Plana]{SpiNNaker1system}
Steve~B. Furber, Francesco Galluppi, Steve Temple, and Luis~A. Plana.
\newblock The spinnaker project.
\newblock \emph{Proceedings of the IEEE}, 102\penalty0 (5):\penalty0 652--665,
  2014.
\newblock \doi{10.1109/JPROC.2014.2304638}.

\bibitem[Gholami et~al.(2021)Gholami, Yao, Kim, Mahoney, and
  Keutzer]{gholami2020}
Amir Gholami, Zhewei Yao, Sehoon Kim, Michael~W Mahoney, and Kurt Keutzer.
\newblock Ai and memory wall.
\newblock \emph{RiseLab Medium Post}, 2021.
\newblock URL \url{https://github.com/amirgholami/ai_and_memory_wall}.

\bibitem[Hammer(2022)]{hammer2022reasoning}
P~Hammer.
\newblock Reasoning-learning systems based on non-axiomatic reasoning system
  theory, vol. 192.
\newblock 2022.

\bibitem[Hooker(2021)]{Hooker2021}
Sara Hooker.
\newblock The hardware lottery.
\newblock \emph{Commun. ACM}, 64\penalty0 (12):\penalty0 58–65, nov 2021.
\newblock ISSN 0001-0782.
\newblock \doi{10.1145/3467017}.
\newblock URL \url{https://doi.org/10.1145/3467017}.

\bibitem[Huang et~al.(2022{\natexlab{a}})Huang, Gerhards, Kreutz, Vogginger,
  Kelber, Scholz, Knobloch, and Mayr]{ref_jennifer_aicas22_2}
Jiaxin Huang, Pascal Gerhards, Felix Kreutz, Bernhard Vogginger, Florian
  Kelber, Daniel Scholz, Klaus Knobloch, and Christian~Georg Mayr.
\newblock Spiking neural network based real-time radar gesture recognition live
  demonstration.
\newblock In \emph{2022 IEEE 4th International Conference on Artificial
  Intelligence Circuits and Systems (AICAS)}, pages 500--500,
  2022{\natexlab{a}}.
\newblock \doi{10.1109/AICAS54282.2022.9869943}.

\bibitem[Huang et~al.(2022{\natexlab{b}})Huang, Vogginger, Gerhards, Kreutz,
  Kelber, Scholz, Knobloch, and Mayr]{ref_jennifer_aicas22_1}
Jiaxin Huang, Bernhard Vogginger, Pascal Gerhards, Felix Kreutz, Florian
  Kelber, Daniel Scholz, Klaus Knobloch, and Christian~Georg Mayr.
\newblock Real-time radar gesture classification with spiking neural network on
  spinnaker 2 prototype.
\newblock In \emph{2022 IEEE 4th International Conference on Artificial
  Intelligence Circuits and Systems (AICAS)}, pages 362--365,
  2022{\natexlab{b}}.
\newblock \doi{10.1109/AICAS54282.2022.9869987}.

\bibitem[Huang et~al.(2023{\natexlab{a}})Huang, Kelber, Vogginger, Liu, Kreutz,
  Gerhards, Scholz, Knobloch, and Mayr]{ref_jennifer_frontiers}
Jiaxin Huang, Florian Kelber, Bernhard Vogginger, Chen Liu, Felix Kreutz,
  Pascal Gerhards, Daniel Scholz, Klaus Knobloch, and Christian~G. Mayr.
\newblock Efficient snn multi-cores mac array acceleration on spinnaker 2.
\newblock \emph{Frontiers in Neuroscience}, 17, 2023{\natexlab{a}}.
\newblock ISSN 1662-453X.
\newblock \doi{10.3389/fnins.2023.1223262}.
\newblock URL
  \url{https://www.frontiersin.org/articles/10.3389/fnins.2023.1223262}.

\bibitem[Huang et~al.(2023{\natexlab{b}})Huang, Kelber, Vogginger, Wu, Kreutz,
  Gerhards, Scholz, Knobloch, and Mayr]{ref_jennifer_AICAS23}
Jiaxin Huang, Florian Kelber, Bernhard Vogginger, Binyi Wu, Felix Kreutz,
  Pascal Gerhards, Daniel Scholz, Klaus Knobloch, and Christian~Georg Mayr.
\newblock Efficient algorithms for accelerating spiking neural networks on mac
  array of spinnaker 2.
\newblock In \emph{2023 IEEE 5th International Conference on Artificial
  Intelligence Circuits and Systems (AICAS)}, pages 1--5, 2023{\natexlab{b}}.
\newblock \doi{10.1109/AICAS57966.2023.10168559}.

\bibitem[Höppner et~al.(2017)Höppner, Yan, Vogginger, Dixius, Partzsch,
  Joshi, Neumärker, Hartmann, Schiefer, Scholze, Ellguth, Cederstroem,
  Eberlein, Mayr, Temple, Plana, Garside, Davison, Lester, and
  Furber]{Hoeppner2017}
Sebastian Höppner, Yexin Yan, Bernhard Vogginger, Andreas Dixius, Johannes
  Partzsch, Prateek Joshi, Felix Neumärker, Stephan Hartmann, Stefan Schiefer,
  Stefan Scholze, Georg Ellguth, Love Cederstroem, Matthias Eberlein, Christian
  Mayr, Steve Temple, Luis Plana, Jim Garside, Simon Davison, David~R. Lester,
  and Steve Furber.
\newblock Live demonstration: Dynamic voltage and frequency scaling for
  neuromorphic many-core systems.
\newblock In \emph{2017 IEEE International Symposium on Circuits and Systems
  (ISCAS)}, pages 1--1, 2017.
\newblock \doi{10.1109/ISCAS.2017.8050396}.

\bibitem[Höppner et~al.(2019)Höppner, Vogginger, Yan, Dixius, Scholze,
  Partzsch, Neumärker, Hartmann, Schiefer, Ellguth, Cederstroem, Plana,
  Garside, Furber, and Mayr]{Hoeppner2019}
Sebastian Höppner, Bernhard Vogginger, Yexin Yan, Andreas Dixius, Stefan
  Scholze, Johannes Partzsch, Felix Neumärker, Stephan Hartmann, Stefan
  Schiefer, Georg Ellguth, Love Cederstroem, Luis~A. Plana, Jim Garside, Steve
  Furber, and Christian Mayr.
\newblock Dynamic power management for neuromorphic many-core systems.
\newblock \emph{IEEE Transactions on Circuits and Systems I: Regular Papers},
  66\penalty0 (8):\penalty0 2973--2986, 2019.
\newblock \doi{10.1109/TCSI.2019.2911898}.

\bibitem[Höppner et~al.(2022)Höppner, Yan, Dixius, Scholze, Partzsch, Stolba,
  Kelber, Vogginger, Neumärker, Ellguth, Hartmann, Schiefer, Hocker, Walter,
  Liu, Garside, Furber, and Mayr]{Hoeppner2022}
Sebastian Höppner, Yexin Yan, Andreas Dixius, Stefan Scholze, Johannes
  Partzsch, Marco Stolba, Florian Kelber, Bernhard Vogginger, Felix Neumärker,
  Georg Ellguth, Stephan Hartmann, Stefan Schiefer, Thomas Hocker, Dennis
  Walter, Genting Liu, Jim Garside, Steve Furber, and Christian Mayr.
\newblock The spinnaker 2 processing element architecture for hybrid digital
  neuromorphic computing, 2022.

\bibitem[Ignjatović et~al.(2022)Ignjatović, Bailey, and
  Bajić]{tenstorrentISSCC2022}
Drago Ignjatović, Daniel~W. Bailey, and Ljubisa Bajić.
\newblock The wormhole ai training processor.
\newblock In \emph{2022 IEEE International Solid- State Circuits Conference
  (ISSCC)}, volume~65, pages 356--358, 2022.
\newblock \doi{10.1109/ISSCC42614.2022.9731633}.

\bibitem[Kappel et~al.(2018)Kappel, Legenstein, Habenschuss, Hsieh, and
  Maass]{Kappel2018}
David Kappel, Robert Legenstein, Stefan Habenschuss, Michael Hsieh, and
  Wolfgang Maass.
\newblock A dynamic connectome supports the emergence of stable computational
  function of neural circuits through reward-based learning.
\newblock \emph{eNeuro}, 5\penalty0 (2), 2018.
\newblock \doi{10.1523/ENEURO.0301-17.2018}.
\newblock URL \url{https://www.eneuro.org/content/5/2/ENEURO.0301-17.2018}.

\bibitem[Kelber et~al.(2020)Kelber, Wu, Vogginger, Partzsch, Liu, Stolba, and
  Mayr]{Kelber2020}
Florian Kelber, Binyi Wu, Bernhard Vogginger, Johannes Partzsch, Chen Liu,
  Marco Stolba, and Christian Mayr.
\newblock Mapping deep neural networks on spinnaker2.
\newblock In \emph{Proceedings of the 2020 Annual Neuro-Inspired Computational
  Elements Workshop}, NICE '20, New York, NY, USA, 2020. Association for
  Computing Machinery.
\newblock ISBN 9781450377188.
\newblock \doi{10.1145/3381755.3381778}.
\newblock URL \url{https://doi.org/10.1145/3381755.3381778}.

\bibitem[Kingma and Ba(2015)]{KingBa15}
Diederik Kingma and Jimmy Ba.
\newblock Adam: A method for stochastic optimization.
\newblock In \emph{International Conference on Learning Representations
  (ICLR)}, San Diega, CA, USA, 2015.

\bibitem[Krizhevsky et~al.(2012)Krizhevsky, Sutskever, and
  Hinton]{Krizhevsky2012}
Alex Krizhevsky, Ilya Sutskever, and Geoffrey~E Hinton.
\newblock Imagenet classification with deep convolutional neural networks.
\newblock In F.~Pereira, C.J. Burges, L.~Bottou, and K.Q. Weinberger, editors,
  \emph{Advances in Neural Information Processing Systems}, volume~25. Curran
  Associates, Inc., 2012.
\newblock URL
  \url{https://proceedings.neurips.cc/paper_files/paper/2012/file/c399862d3b9d6b76c8436e924a68c45b-Paper.pdf}.

\bibitem[Liu et~al.(2018)Liu, Bellec, Vogginger, Kappel, Partzsch, Neumärker,
  Höppner, Maass, Furber, Legenstein, and Mayr]{Liu2018}
Chen Liu, Guillaume Bellec, Bernhard Vogginger, David Kappel, Johannes
  Partzsch, Felix Neumärker, Sebastian Höppner, Wolfgang Maass, Steve~B.
  Furber, Robert Legenstein, and Christian~G. Mayr.
\newblock Memory-efficient deep learning on a spinnaker 2 prototype.
\newblock \emph{Frontiers in Neuroscience}, 12, 2018.
\newblock ISSN 1662-453X.
\newblock \doi{10.3389/fnins.2018.00840}.
\newblock URL
  \url{https://www.frontiersin.org/articles/10.3389/fnins.2018.00840}.

\bibitem[Painkras et~al.(2013)Painkras, Plana, Garside, Temple, Galluppi,
  Patterson, Lester, Brown, and Furber]{SpiNNaker1}
Eustace Painkras, Luis~A. Plana, Jim Garside, Steve Temple, Francesco Galluppi,
  Cameron Patterson, David~R. Lester, Andrew~D. Brown, and Steve~B. Furber.
\newblock Spinnaker: A 1-w 18-core system-on-chip for massively-parallel neural
  network simulation.
\newblock \emph{IEEE Journal of Solid-State Circuits}, 48\penalty0
  (8):\penalty0 1943--1953, 2013.
\newblock \doi{10.1109/JSSC.2013.2259038}.

\bibitem[Pedersen et~al.(2023)Pedersen, Abreu, Jobst, Lenz, Fra, Bauer, Muir,
  Zhou, Vogginger, Heckel, Urgese, Shankar, Stewart, Eshraghian, and
  Sheik]{pedersen2023neuromorphic}
Jens~E. Pedersen, Steven Abreu, Matthias Jobst, Gregor Lenz, Vittorio Fra,
  Felix~C. Bauer, Dylan~R. Muir, Peng Zhou, Bernhard Vogginger, Kade Heckel,
  Gianvito Urgese, Sadasivan Shankar, Terrence~C. Stewart, Jason~K. Eshraghian,
  and Sadique Sheik.
\newblock Neuromorphic intermediate representation: A unified instruction set
  for interoperable brain-inspired computing, 2023.

\bibitem[Pehle and Pedersen(2021)]{norse2021}
Christian Pehle and Jens~Egholm Pedersen.
\newblock {Norse - A deep learning library for spiking neural networks},
  January 2021.
\newblock URL \url{https://doi.org/10.5281/zenodo.4422025}.
\newblock Documentation: https://norse.ai/docs/.

\bibitem[Pehle et~al.(2022)Pehle, Billaudelle, Cramer, Kaiser, Schreiber,
  Stradmann, Weis, Leibfried, M{\"u}ller, and Schemmel]{pehle2022brainscales}
Christian Pehle, Sebastian Billaudelle, Benjamin Cramer, Jakob Kaiser,
  Korbinian Schreiber, Yannik Stradmann, Johannes Weis, Aron Leibfried, Eric
  M{\"u}ller, and Johannes Schemmel.
\newblock The brainscales-2 accelerated neuromorphic system with hybrid
  plasticity.
\newblock \emph{Frontiers in Neuroscience}, 16:\penalty0 795876, 2022.

\bibitem[Peng et~al.(2023)Peng, Alcaide, Anthony, Albalak, Arcadinho, Cao,
  Cheng, Chung, Grella, GV, He, Hou, Kazienko, Kocon, Kong, Koptyra, Lau,
  Mantri, Mom, Saito, Tang, Wang, Wind, Wozniak, Zhang, Zhang, Zhao, Zhou, Zhu,
  and Zhu]{peng2023rwkv}
Bo~Peng, Eric Alcaide, Quentin Anthony, Alon Albalak, Samuel Arcadinho, Huanqi
  Cao, Xin Cheng, Michael Chung, Matteo Grella, Kranthi~Kiran GV, Xuzheng He,
  Haowen Hou, Przemyslaw Kazienko, Jan Kocon, Jiaming Kong, Bartlomiej Koptyra,
  Hayden Lau, Krishna Sri~Ipsit Mantri, Ferdinand Mom, Atsushi Saito, Xiangru
  Tang, Bolun Wang, Johan~S. Wind, Stansilaw Wozniak, Ruichong Zhang, Zhenyuan
  Zhang, Qihang Zhao, Peng Zhou, Jian Zhu, and Rui-Jie Zhu.
\newblock Rwkv: Reinventing rnns for the transformer era, 2023.

\bibitem[Prabhakar et~al.(2022)Prabhakar, Jairath, and
  Shin]{sambanovaISSCC2022}
Raghu Prabhakar, Sumti Jairath, and Jinuk~Luke Shin.
\newblock Sambanova sn10 rdu: A 7nm dataflow architecture to accelerate
  software 2.0.
\newblock In \emph{2022 IEEE International Solid- State Circuits Conference
  (ISSCC)}, volume~65, pages 350--352, 2022.
\newblock \doi{10.1109/ISSCC42614.2022.9731612}.

\bibitem[Rostami et~al.(2022)Rostami, Vogginger, Yan, and Mayr]{AmirEprop}
Amirhossein Rostami, Bernhard Vogginger, Yexin Yan, and Christian~G. Mayr.
\newblock E-prop on spinnaker 2: Exploring online learning in spiking rnns on
  neuromorphic hardware.
\newblock \emph{Frontiers in Neuroscience}, 16, 2022.
\newblock ISSN 1662-453X.
\newblock \doi{10.3389/fnins.2022.1018006}.
\newblock URL
  \url{https://www.frontiersin.org/articles/10.3389/fnins.2022.1018006}.

\bibitem[Rowley et~al.(2019)Rowley, Brenninkmeijer, Davidson, Fellows, Gait,
  Lester, Plana, Rhodes, Stokes, and Furber]{rowley2019spinntools}
Andrew~GD Rowley, Christian Brenninkmeijer, Simon Davidson, Donal Fellows,
  Andrew Gait, David~R Lester, Luis~A Plana, Oliver Rhodes, Alan~B Stokes, and
  Steve~B Furber.
\newblock Spinntools: the execution engine for the spinnaker platform.
\newblock \emph{Frontiers in neuroscience}, 13:\penalty0 231, 2019.

\bibitem[Rueckauer et~al.(2017)Rueckauer, Lungu, Hu, Pfeiffer, and
  Liu]{ref_snntoolbox}
Bodo Rueckauer, Iulia-Alexandra Lungu, Yuhuang Hu, Michael Pfeiffer, and
  Shih-Chii Liu.
\newblock Conversion of continuous-valued deep networks to efficient
  event-driven networks for image classification.
\newblock \emph{Frontiers in Neuroscience}, 11, 2017.
\newblock ISSN 1662-453X.
\newblock \doi{10.3389/fnins.2017.00682}.
\newblock URL
  \url{https://www.frontiersin.org/articles/10.3389/fnins.2017.00682}.

\bibitem[Sevilla et~al.(2022)Sevilla, Heim, Ho, Besiroglu, Hobbhahn, and
  Villalobos]{Sevilla2022}
Jaime Sevilla, Lennart Heim, Anson Ho, Tamay Besiroglu, Marius Hobbhahn, and
  Pablo Villalobos.
\newblock Compute trends across three eras of machine learning.
\newblock In \emph{2022 International Joint Conference on Neural Networks
  (IJCNN)}, pages 1--8, 2022.
\newblock \doi{10.1109/IJCNN55064.2022.9891914}.

\bibitem[Subramoney et~al.(2023)Subramoney, Nazeer, Sch{\"o}ne, Mayr, and
  Kappel]{Subramoney2023}
Anand Subramoney, Khaleelulla~Khan Nazeer, Mark Sch{\"o}ne, Christian Mayr, and
  David Kappel.
\newblock Efficient recurrent architectures through activity sparsity and
  sparse back-propagation through time.
\newblock In \emph{The Eleventh International Conference on Learning
  Representations}, 2023.
\newblock URL \url{https://openreview.net/forum?id=lJdOlWg8td}.

\bibitem[Sun et~al.(2023)Sun, Dong, Huang, Ma, Xia, Xue, Wang, and
  Wei]{sun2023retentive}
Yutao Sun, Li~Dong, Shaohan Huang, Shuming Ma, Yuqing Xia, Jilong Xue, Jianyong
  Wang, and Furu Wei.
\newblock Retentive network: A successor to transformer for large language
  models, 2023.

\bibitem[Taherkhani et~al.(2020)Taherkhani, Belatreche, Li, Cosma, Maguire, and
  McGinnity]{Taherkhani2020}
Aboozar Taherkhani, Ammar Belatreche, Yuhua Li, Georgina Cosma, Liam~P.
  Maguire, and T.M. McGinnity.
\newblock A review of learning in biologically plausible spiking neural
  networks.
\newblock \emph{Neural Networks}, 122:\penalty0 253--272, 2020.
\newblock ISSN 0893-6080.
\newblock \doi{https://doi.org/10.1016/j.neunet.2019.09.036}.
\newblock URL
  \url{https://www.sciencedirect.com/science/article/pii/S0893608019303181}.

\bibitem[Tavanaei et~al.(2019)Tavanaei, Ghodrati, Kheradpisheh, Masquelier, and
  Maida]{Tavanaei2019}
Amirhossein Tavanaei, Masoud Ghodrati, Saeed~Reza Kheradpisheh, Timothée
  Masquelier, and Anthony Maida.
\newblock Deep learning in spiking neural networks.
\newblock \emph{Neural Networks}, 111:\penalty0 47--63, 2019.
\newblock ISSN 0893-6080.
\newblock \doi{https://doi.org/10.1016/j.neunet.2018.12.002}.
\newblock URL
  \url{https://www.sciencedirect.com/science/article/pii/S0893608018303332}.

\bibitem[Thompson et~al.(2022)Thompson, Greenewald, Lee, and
  Manso]{thompson2022computational}
Neil~C. Thompson, Kristjan Greenewald, Keeheon Lee, and Gabriel~F. Manso.
\newblock The computational limits of deep learning, 2022.

\bibitem[Vogginger et~al.(2023)Vogginger, Kelber, Jobst, Yan, Gerhards, Weih,
  and Akl]{vogginger2023pyspinnaker2}
Bernhard Vogginger, Florian Kelber, Matthias Jobst, Yexin Yan, Pascal Gerhards,
  Martin Weih, and Mahmoud Akl.
\newblock py-spinnaker2, November 2023.
\newblock URL \url{https://doi.org/10.5281/zenodo.10202110}.

\bibitem[Wo{\'{z}}niak et~al.(2020)Wo{\'{z}}niak, Pantazi, Bohnstingl, and
  Eleftheriou]{Wozniak2020}
Stanis{\l}aw Wo{\'{z}}niak, Angeliki Pantazi, Thomas Bohnstingl, and Evangelos
  Eleftheriou.
\newblock Deep learning incorporating biologically inspired neural dynamics and
  in-memory computing.
\newblock \emph{Nature Machine Intelligence}, 2\penalty0 (6):\penalty0
  325--336, Jun 2020.
\newblock ISSN 2522-5839.
\newblock \doi{10.1038/s42256-020-0187-0}.
\newblock URL \url{https://doi.org/10.1038/s42256-020-0187-0}.

\bibitem[Wunderlich and Pehle(2021)]{Wunderlich2021}
Timo~C. Wunderlich and Christian Pehle.
\newblock Event-based backpropagation can compute exact gradients for spiking
  neural networks.
\newblock \emph{Scientific Reports}, 11\penalty0 (1):\penalty0 12829, Jun 2021.
\newblock ISSN 2045-2322.
\newblock \doi{10.1038/s41598-021-91786-z}.

\bibitem[Yamazaki et~al.(2022)Yamazaki, Vo-Ho, Bulsara, and Le]{Yamazaki2022}
Kashu Yamazaki, Viet-Khoa Vo-Ho, Darshan Bulsara, and Ngan Le.
\newblock Spiking neural networks and their applications: A review.
\newblock \emph{Brain Sciences}, 12\penalty0 (7), 2022.
\newblock ISSN 2076-3425.
\newblock \doi{10.3390/brainsci12070863}.
\newblock URL \url{https://www.mdpi.com/2076-3425/12/7/863}.

\bibitem[Yan(2022)]{Yexin2022}
Yexin Yan.
\newblock \emph{Implementation of bioinspired algorithms on the neuromorphic
  VLSI system SpiNNaker 2}.
\newblock PhD thesis, TU Dresden, 2022.

\bibitem[Yan et~al.(2019)Yan, Kappel, Neumärker, Partzsch, Vogginger,
  Höppner, Furber, Maass, Legenstein, and Mayr]{Yan2019}
Yexin Yan, David Kappel, Felix Neumärker, Johannes Partzsch, Bernhard
  Vogginger, Sebastian Höppner, Steve Furber, Wolfgang Maass, Robert
  Legenstein, and Christian Mayr.
\newblock Efficient reward-based structural plasticity on a spinnaker 2
  prototype.
\newblock \emph{IEEE Transactions on Biomedical Circuits and Systems},
  13\penalty0 (3):\penalty0 579--591, 2019.
\newblock \doi{10.1109/TBCAS.2019.2906401}.

\end{thebibliography}

\end{document}